\documentclass{article}

\usepackage{arxiv}

\usepackage[utf8]{inputenc} 
\usepackage[T1]{fontenc}    
\usepackage{hyperref}       
\usepackage{url}            
\usepackage{booktabs}       
\usepackage{amsfonts}       
\usepackage{nicefrac}       
\usepackage{microtype}      
\usepackage{lipsum}
\usepackage{graphicx}
\graphicspath{ {./images/} }
\usepackage{amsmath} 
\usepackage{amssymb}

\title{Optimal Synchronization Control for Heterogeneous Multi-Agent Systems: Online Adaptive Learning Solutions}

\author{
 Yuanqiang Zhou \\
  Department of Automation\\
  Shanghai Jiao Tong University\\
  Shanghai, China 200240 \\
  \texttt{zhouyuanq@gmail.com} \\
   \And
 Dewei Li \\
 Department of Automation\\
 Shanghai Jiao Tong University\\
 Shanghai, China 200240 \\
 \texttt{dwli@sjtu.edu.cn} \\
  \And
 Furong Gao \\
 Department of Chemical and Biological Engineering\\
 The Hong Kong University of Science and Technology\\
 Clear Water Bay, Hong Kong\\
  \texttt{kefgao@ust.hk} \\
}

\begin{document}
	
	\maketitle
	
	\begin{abstract}
		This paper presents an online adaptive learning solution to optimal synchronization control problem of  heterogeneous multi-agent systems via a novel distributed policy iteration approach. 
		For the leader-follower multi-agents, the dynamics of all the followers are heterogeneous with leader disturbance. 
		To make the output of each follower synchronize with the leader's output, we propose a synchronization control protocol where the stability conditions for selecting the feedback gains are given. 
		Then, with a minimization of the output tracking errors, we optimize the feedback gains for the synchronization control protocol and the unique solutions for those feedback gains are learned via a novel distributed policy iteration approach. The proposed online adaptive learning solution results in the optimal control solution. Finally, an illustrative numerical example is provided to show the effectiveness of our approaches. 
		
		\keywords{Adaptive control, synchronization control, policy iteration (PI), distributed control, multi-agent system (MAS)}
	\end{abstract}
	
	\section{Introduction}  \label{sec1}
	Synchronization control of multi-agent systems has attracted notable attention in recent years due to its wide applications in electrical, mechanical, and biological systems; see \cite{Ren2008,Huang2004,Yang2017,Su2012,Xu2019synchronization,xu2019event}. However, a general assumption on this topic is that the systems are homogeneous without considering influence of the external disturbances \cite{you2017distributed,Li2015}. This motivates us to investigate the synchronization control of the heterogeneous disturbed multi-agent systems with assured output synchronization and optimality. 
	
	Recently, some preliminary results have been proposed on synchronization control for various classes of systems and time-varying networks, while the majority of existing results have focused on the controller synthesis problems that enforce a set of general linear systems to reach consensus asymptotically, see e.g. \cite{Meng2018,Cai2017}. The authors in \cite{Abdessameud2018} considered the high-order MASs described by multiple integrator dynamic under general directed graphs; the authors in \cite{Yang2017} investigated the distributed tracking control problem for a class of Euler-Lagrange MASs; the authors in \cite{Liuzza2016} investigated the problem of event-triggered control for the synchronization of networks of nonlinear dynamical agents; the authors in \cite{Chen2014} considered the consensus control of nonlinear MASs with time-delay states. In \cite{Sadikhov2017}, the consensus problem for a group of agent robots with a connected, undirected, and time-invariant communication graph topology in the face of uncertain inter-agent measurement data was addressed. In \cite{Zhao2019}, an distributed adaptive approach was developed for  time-varying formation problem of linear MASs under directed topologies. 
	However, most of the aforementioned references consider the synchronization control problems, but not consider the optimal synchronization control problems. Although the authors in \cite{Jiang2019} considered the optimal output regulation, it is used for one discrete-time system under disturbance to synchronize with the leader, not for the heterogeneous MAS case. The authors in\cite{Qin2019} considered the optimal synchronization control of multiagent systems, but it is a homogeneous MAS case. The authors in \cite{vamvoudakis2012multi} proposed the online adaptive learning solution for synchronization control of a set of general linear systems, but they do not consider the disturbance for each follower of the multi-agent systems. Thus, it is still a challenge to find optimal synchronization control solutions when there are some disturbances in the dynamics of the followers. 
	
	In this paper, we provide the online adaptive learning solutions for optimal synchronization control of heterogeneous MASs via a novel distributed policy iteration approach.  Policy iteration (PI) approach is a two-step iteration method for finding the optimal solutions \cite{Sutton1998,Bertsekas2008}. Since for the leader-follower multi-agents, the dynamics of all the followers are heterogeneous with leader disturbance, we develop a synchronization control protocol 
	which transform the disturbed multi-agent systems into the general representation form, then utilize a novel distributed policy iteration approach to online learn the optimal solutions.  
	%
	%
	The main contributions of this paper are threefold. 
	First, we develop the adaptive optimal control architectures which guarantee the output synchronization, while minimizing tracking errors and rejecting disturbance. 
	Next,  a distributed synchronization control protocol is proposed and the optimal feedback gains of the distributed protocol are found via online distributed PI technique by minimizing the output tracking errors. 
	Third, it is worth emphasizing that, different from the results presented in \cite{vamvoudakis2012multi,Meng2018}, this paper investigates the adaptive optimal synchronization control problems where the system's (or, the follower's) model has a direct feedthrough, that is, $D_i \neq 0$ in (3) and a disturbance term from their common leader. The direct feedthrough term will result in a cross-product from in the quadratic performance function and the leader disturbance term will make designing the unique feedback solution difficult.

	The rest of this paper is organized as follows. In Section \ref{sec2}, we introduce the considered systems and the necessary assumptions. 
	In Section \ref{sec3}, we propose a distributed synchronization control protocol for the heterogeneous MASs. 
	In Section \ref{sec4}, we develop the online adaptive learning solutions for optimal synchronization control of heterogeneous MASs via a novel distributed policy iteration approach.
	Convergence of the proposed algorithms and stability of the closed-loop systems are analyzed, respectively. 
	In Section \ref{sec5}, an illustrative numerical example is provided to validate our design. Finally, we draw concluding remarks in Section \ref{sec6}.

	The notation used in this paper is fairly standard. Specifically, $\mathbb{R}$ denotes the set of real numbers, $|\cdot|$ represents the Euclidean norm for a vector and the induced norm for a matrix. 
	%
	For a matrix $A \in \mathbb{R}^{n \times n}$, $A \succ 0$ and $A \succeq 0$ denote that $A$ is positive definite and positive semidefinite, respectively; $\lambda_M(A)$ and $\lambda_m(A)$ denote the maximum and the minimum eigenvalue of the real symmetric matrix $A$, respectively; $\sigma(A)$ denotes the complex spectrum of $A$. 
	For a vector $x \in \mathbb{R}^n$, ${| x |^2_P}$ denotes the quadratic form $ x^\mathrm{T}Px$ for a real symmetric and positive semidefinite matrix $P$.

	\section{Problem Formulation} \label{sec2}
	\subsection{System Description}
	Consider a class of nonidentical $N+1$ multi-agent systems consisting of a leader and $N$ followers, indexed by agent $0$ and agents $1, \ldots, N$, respectively. The dynamics of the agent $i$, $i =1, \ldots, N$ are given by
	\begin{align}   \label{eq:1}
		\dot{x}_{i}(t) & = A_{i}x_{i} (t) +B_{i}u_{i}(t)+ \Delta_i(t),  \quad  x_{i}(0)  =x_{i0}, \quad t \ge 0,  \\
		y_{i}(t) & = C_{i}x_{i}(t)+D_{i}u_{i}(t),  
	\end{align}
	where $x_{i} \in \mathbb{R}^{n_{i}}$ and $u_{i}\in\mathbb{R}^{m_{i}}$ are the state and the input of the $i$th agent, respectively. 
	$x_{i0}$, $i=1, \ldots, N$, represents the initial condition of the agent $i$. 
	
	In some practical applications, the leader acts as a reference generator that provides desired trajectory for the following agents to track. So, the dynamics of the agent 0 are described
	by 
	\begin{align}
		\dot{w}(t) & =Sw(t), \quad w(0) =w_{0},   \quad t \ge 0,  \label{eq:2} \\
		y_{ri}(t) & = F_{i} w(t) \in {\mathbb{R}}^{p_{i}} 
	\end{align}
	where $w \in {\mathbb{R}}^{q}$ is the state of the leader and $w_{0}$ denotes the initial condition of the leader. 
	
	In addition, we assume that the agent $0$ generates the disturbance $\Delta_i(t)$ in \eqref{eq:1} as
	\begin{align} \label{eq:3}
		\Delta_i(t) = E_{i}w(t). 
	\end{align}
	In this case, the output error for each agent $i$ is defined as
	\begin{align} \label{eq:4}
		e_{i}(t) & \buildrel \vartriangle \over = y_{i}(t) - y_{ri}(t) 
	\end{align}
	
	
	For the disturbed MASs given by \eqref{eq:1}--\eqref{eq:3} , we give the following assumptions.

	\textbf{Assumption~1.}    $(A_{i}, C_{i})$ is observable; $D_{i}^\mathrm{T}D_{i}$ is invertible, for all $i = 1, \ldots, N$.  
	
	\textbf{Assumption~2.}    There exists $K_i$ such that, for each $i = 1, \ldots, N$, $A_{i} - B_{i} K_i$ is Hurwitz.  
	
	\textbf{Assumption~3.}  The real parts of eigenvalues of $S$ are non-negative.   
	
	\textbf{Assumption~4.}  For all $\lambda \in \sigma(S)$, where $\sigma(S)$ denotes the complex spectrum of $S$, $\mathrm{rank} \begin{bmatrix} A_{i}-\lambda I_{n_{i}} & B_{i}\\ C_{i} & D_{i} \end{bmatrix}  = n_{i} + m_{i}$,  $i=1, \ldots, N$. 
	
	Assumptions 1-4 are reasonable and necessary. Similar assumptions can be found in \cite{Huang2004,Meng2018,Cai2017,Odekunle2020,Zhang2017} for solving synchronization control problems. 
	
	\subsection{Communication Graph}
	
	This paper considers a scenario in which $w$ can only be accessed by some of the $N$ followers and the leader only disturbs those followers who has information interacted with it. 
	%
	The communication network topology among those followers is denoted as a directed graph $\mathcal{G}=(\mathcal{V},\mathcal{E}, \mathcal{A})$, where $\mathcal{V}=\{v_{0},v_{1}, \ldots, v_{N}\}$ is the node set with $v_{0}$ denoting the leader modeled via the equation \eqref{eq:2} and $v_{i}$, $i=1, \ldots, N$, representing each agent $i$  by \eqref{eq:1} being identified as followers.  $\mathcal{E}=\left\{ \rho_{ij}=\left(v_{j},v_{i}\right)\right\}  \subset \mathcal{V}\times\mathcal{V}$ denotes the edge set of $\mathcal{G}$,  
	$(v_{i},v_{j})\in\mathcal{E}$ means that there exists a direct path form node $v_{i}$ to node $v_{j}$ and node $v_{j}$ can receive information from node $v_{i}$, but not vice ersa. A path from $v_{i}$ to $v_{j}$ in the digraph $\mathcal{G}$ is a sequence of edges $(v_{i},v_{i_{1}}),(v_{i_{1}},v_{i_{2}}),\ldots,(v_{i_{m}},v_{j})$ with distinct nodes $i_{k},k=1,...,m$, then $v_{i}$ is said to be reachable from node $v_{j}$. If $v_{i}=v_{j}$, then the path is called a loop. If a node (called the root) is reachable from every other node of $\mathcal{G}$, then the graph is referred to as a spanning tree. 
	$\mathcal{A}=[\rho_{ij}] \in \mathbb{R}^{(N+1)\times (N+1)}$ denotes the weighted adjacency matrix with $\rho_{ij} =1$, if $\rho_{ij}\in\mathcal{E}$, and $\rho_{ij}=0$, otherwise. In this case, we let 
	\begin{equation*}
		\mathcal{L}=\mathcal{D}-\mathcal{A}, 
	\end{equation*}
	denote the Laplacian matrix of  $\mathcal{G}$, where 
	\begin{equation*}
		\mathcal{D}=\mathrm{diag} \left[ d_{0},d_{1}, \ldots, d_{N} \right] \in\mathbb{R}^{(N+1)\times(N+1)}, \quad  d_{i}=\sum_{j = 0}^N \rho_{ij}, \quad  i=0, \ldots, N. 
	\end{equation*}
	is the in-degree matrix of graph $\mathcal{G}$. Although the leader $v_{0}$ can be reachable by the other followers, it can not access information from the other followers. In this case, the Laplacian matrix of $\mathcal{G}$ has the following form
	\begin{align} \label{eq:5}
		\mathcal{L}=\mathcal{D}-\mathcal{A}=\left[\begin{array}{cc}
			0 & \textbf{0}\\
			-\mathcal{A}_{0}\textbf{1}_{N} & \mathcal{H}
		\end{array}\right], 
	\end{align}
	where $\mathcal{H} \buildrel \vartriangle \over = \mathcal{A}_{0}+\mathcal{L}_{s}$, $\mathcal{A}_{0}\buildrel \vartriangle \over = \mathrm{diag} \left[ \rho_{10}, \ldots, \rho_{N0} \right]$, and $\mathcal{L}_{s}$ is the Laplacian matrix of subgraph $\mathcal{G}_{s}$ with node $\{v_{1},\ldots, v_{N}\}$. For more details, one can refer to \cite{Ren2008,Li2015}.
	

	\textbf{Assumption~5.}  
	The communication topology $\mathcal{G}$ contains no loop and has a directed spanning tree with $v_{0}$ as its root.

	Following the results in \cite{Huang2004,Su2012},  Assumption 5 implies that zero is a eigenvalue of $\mathcal{L}$ in \eqref{eq:5} with $\mathbb{1}$ as a right eigenvector and all nonzero eigenvalues have positive real parts. 
	
	\emph{Problem~1:} Optimal  synchronization control problem is solved for the disturbed MASs given by \eqref{eq:1}--\eqref{eq:3}  in the communication network $\mathcal{G}$, if a distributed control protocol can be designed such that the following conditions hold: 
	\begin{itemize}
		\item \emph{Output synchronization}: the error given by \eqref{eq:3} satisfies
		\begin{align}  \label{eq:3-1}
			\mathrm{lim}_{t\to\infty} e_{i}(t)=0
		\end{align} 
		for all $i = 1, \ldots, N$.
		\item \emph{Output error minimization}: the control input $u_i$ for each agent $i$ satisfies
		\begin{align}  \label{eq:3-2}
			u_i(t) = \mathop {\arg \min} J_{i}(X_{i0}) 
		\end{align}
		where 
		\begin{align}  \label{eq:31}
			J_{i}(X_{i0}) =\int_{0}^{\infty}e_{i}^\mathrm{T} (t) e_{i}(t) \mathrm{d}t.
		\end{align}
	\end{itemize}
	
	The  main objective of this paper is to design a control protocol such that the proposed \emph{Problem~1} can be solved in the communication network $\mathcal{G}$ under Assumptions 1-5. In the following sections, we propose a distributed control protocol which realizes the output synchronization and then, based on this initial protocol, we learn the unique optimal feedback solution via online distributed policy iteration technique by minimizing the output tracking errors.
	
	\section{Synchronization Control of Heterogeneous MASs}  \label{sec3}
	Using Assumptions 1-5, several control protocols have been proposed for the MASs given by (1) and (2) without considering the external disturbances, see e.g.,  \cite{Li2015,Cai2017}. For our disturbed MASs given by (1) and (2),  
	a distributed compensator based on the leading system \eqref{eq:2} is developed as
	\begin{align}\label{eq:6}
		\dot{\xi}_{i} (t)  & = S\xi_{i}(t)+\alpha_{i}\Bigg[ \sum_{j=0}^{N}\rho_{ij}(\xi_{i}(t)-\xi_{j}(t)) \Bigg],   \notag \\
		& \qquad \qquad \qquad \qquad  \quad \quad \xi_{i}(0) = \xi_{i0},  \quad t \ge 0, 
	\end{align}
	where $i = 0, 1, \ldots, N$, $\xi_{0}(t)= w(t)$, $t \ge 0$, and $\alpha_{i} \in\mathbb{R}$. 
	Then we have the following lemma which show asymptotic convergence of the compensator \eqref{eq:6} to the leader \eqref{eq:2}. 

	\textbf{Lemma~1.}  
	Consider the heterogeneous MASs in the communication network $\mathcal{G}$ and assume that Assumptions 1-5 hold. Choose $\alpha_{i}\in\mathbb{R}$ such that, for $i =1, \ldots, N$, 
	\begin{align}\label{eq:7}
		\alpha_{i}d_{i}+\lambda_{M}(S) < 0. 
	\end{align}
	Then for $\varsigma_{i}(t) \buildrel \vartriangle \over = \xi_{i} (t) - w(t)$, $t \ge 0$, $\mathrm{lim}_{t\to\infty} |\varsigma_{i}(t)| = 0$. 
	
	\emph{Proof.}
		%
		According to \eqref{eq:2} and \eqref{eq:6}, we obtain
		\begin{align} \label{eq:8}
			\dot{\varsigma}_{i}(t) 
			& = S \varsigma_{i}(t)+\alpha_{i}\left[ \sum_{j=1}^{N}\rho_{ij} (\xi_{i}(t)-\xi_{j}(t))+ \rho_{i0} \varsigma_{i}(t)\right] \notag\\ 
			& =\left(S+\alpha_{i} \rho_{i0} I_{q}\right)\varsigma_{i}(t)+\alpha_{i}(\mathcal{L}_{s}\otimes I_{q})\varsigma(t). 
		\end{align}
		
		Next, we let $\varsigma(t) =\left[\varsigma_{1}^\mathrm{T}(t), \ldots, \varsigma_{N}^\mathrm{T}(t)\right]^\mathrm{T}$, $t\ge 0$, and then, putting \eqref{eq:8} in a compact form yields 
		\begin{align}\label{eq:9}
			\dot{\varsigma}(t) & =\left(I_{N}\otimes S+\Lambda\mathcal{A}_{0}\otimes I_{q}+\Lambda\mathcal{L}_{s}\otimes I_{q}\right)\varsigma(t) \notag \\
			& =\left(I_{N}\otimes S+\Lambda\mathcal{H}\otimes I_{q}\right) \varsigma(t), \quad t \ge 0, 
		\end{align}
		where $\Lambda \buildrel \vartriangle \over = \mathrm{diag}\left[ \alpha_{1}, \ldots, \alpha_{N}\right]$. Assumption 3 indicates that all the eigenvalues of $\mathcal{H}$ have positive real parts, that is, $d_{i}>0$, $i = 1, \ldots, N$.  In addition, we note that the eigenvalues of $\Lambda\mathcal{H}$ are $\lambda_{i}(\Lambda\mathcal{\mathcal{H}})=$ $\alpha_{i}d_{i}$, $i = 1, \ldots, N$. Thus the eigenvalues of $\mathcal{M}_{i}  \buildrel \vartriangle \over = I_{N}\otimes S+\alpha_{i}\mathcal{H}\otimes I_{q}$ are $\lambda_{j}(S)+\alpha_{i}d_{i}$, $i = 1, \ldots, N$, $j = 1, \ldots, q$. 
		Since $\alpha_{i}$ satisfies \eqref{eq:7} for all $i =1, \ldots, N$, $\varsigma(t)$ in \eqref{eq:9} is asymptotically stable, which implies the asymptotic stability of $\varsigma_{i}(t)$, $t \ge 0$. Therefore, $\mathrm{lim}_{t\to\infty} |\varsigma_{i}(t)| = \mathrm{lim}_{t\to\infty} |\xi_{i}(t) - w(t)| =0$ for $i = 1, \ldots, N$, which completes the proof. 

	According to Lemma 1 and for a given real constant $r > 0$, we design $\alpha_{i}$ to satisfy
	\begin{align}
		\alpha_{i} d_i = -\lambda_{M}(S) - r,  \label{eq:101}
	\end{align}
	for $i =1, \ldots, N$, which indicates that $ \alpha_{i} d_i + \lambda_{M}(S) = - r <  0$, $i =1, \ldots, N$.
	
	Next, based on \eqref{eq:6}, we  design the distributed control protocols for the disturbed MASs given by (1) and (2) as
	\begin{align}
		\dot{\zeta}_{i}(t) & = \Big[ S-(\lambda_{M}(S)+r)I_{q} \Big] {\zeta}_{i}(t), \notag \\
		& \qquad \qquad \qquad \qquad  \quad {\zeta}_{i}(0) = {\zeta}_{i0}, \quad t \ge 0,  \label{eq:10} \\
		u_{i}(t) & = - K_{i,1} x_{i}(t) - K_{i,2} \xi_{i}(t) - K_{i,3} \zeta_{i}(t), \label{eq:11}
	\end{align}
	for $i = 1, \ldots, N$, where $\xi_{i} \in \mathbb{R}^q$, $\zeta_{i} \in\mathbb{R}^{q}$, $\alpha_{i}\in\mathbb{R}$, ${\zeta}_{i0}$ is the initial condition to be given later, and the matrices $K_{1i} \in \mathbb{R}^{m_i \times n_i}$, $K_{i,2} \in \mathbb{R}^{m_i \times q}$ and $K_{i,3} \in \mathbb{R}^{m_i \times q}$ are the control gains to be designed later. 
	
	For the distributed control protocol \eqref{eq:11}, $\xi_{i}$ and $\zeta_{i}$ in \eqref{eq:6} and \eqref{eq:10}, are the augmented states. 
	In particular, $\xi_{i}$ is transmitted over the communication network $\mathcal{G}$, while $\zeta_{i}$ is used locally by agent $i$ only.  
	Next, we have the following theorem which ensures the output synchronization for the heterogeneous MASs given by \eqref{eq:1}-\eqref{eq:4} under the distributed controllers \eqref{eq:6}, \eqref{eq:10}, and \eqref{eq:11}.

	
	\textbf{Theorem~1. }
	Consider the heterogeneous MASs in the communication network with graph $\mathcal{G}$, assume that Assumptions 1-5 hold, and given a real constant $r >0$. Let \eqref{eq:17} hold for \eqref{eq:10}  and choose $\alpha_{i}\in\mathbb{R}$, $K_{i,1}\in\mathbb{R}^{m_{i}\times n_{i}}$, $K_{i,2}\in\mathbb{R}^{m_{i} \times q}$, and $K_{i,3}\in\mathbb{R}^{m_{i}\times q}$, such that 
	\begin{align}
		& \alpha_{i}d_{i}+\lambda_{M}(S)  = - r ,  \label{eq:21-1}\\
		& A_{i} - B_{i}K_{i,1} ~\mathit{is} ~\mathit{Hurwitz},  \label{eq:21-2} \\
		&K_{i,1}\Pi_{i}+K_{i,2}+\Gamma_{i} = 0,  \label{eq:21-3}
	\end{align}
	where the pairs $(\Pi_{i}, \Gamma_{i})$ are the solutions to the following regulator equations
	\begin{align}
		\Pi_{i}S & =A_{i}\Pi_{i}+B_{i}\Gamma_{i}+E_{i}, \label{eq:22-1}\\
		0 & =C_{i}\Pi_{i}+D_{i}\Gamma_{i}-F_{i}. \label{eq:22-2}
	\end{align}
	Then, for any initial conditions $x_{i0}\in\mathbb{R}^{n_{i}}$ and $w_{0}\in\mathbb{R}^{q}$, the tracking errors in \eqref{eq:4} satisfy
	\begin{align*}
		\mathrm{lim}_{t\to\infty} e_{i}(t)=0, \quad i = 1, \ldots, N.
	\end{align*}

	\emph{Proof. }
		Using Assumption 4 and similar arguments as in \cite{Huang2004}, we have that for each agent $i= 1, \ldots, N$, there exist unique matrix pairs ($\Pi_{i}, \Gamma_{i}$) for solving \eqref{eq:22-1} and \eqref{eq:22-2}. 
		Next, note that although $\xi_{i}(t)$, $t \ge 0$, in \eqref{eq:6} is transmitted over the communication network with graph $\mathcal{G}$, because $w(t)$, $t \ge 0$,  is not available for all agent $i = 1, \ldots, N$, the signal $\varsigma_{i}(t)$, $t \ge 0$, can not be obtained directly for all agent $i = 1, \ldots, N$. Thus, according to \eqref{eq:9} and \eqref{eq:10}, 
		using \eqref{eq:9} and \eqref{eq:10}, we construct a transformation for the MAS given by (1) and (2) as
		\begin{align*}
			\zeta(t) = (U\otimes I_{q}) \varsigma(t), \quad t \ge 0, 
		\end{align*}
		where ${\zeta} \buildrel \vartriangle \over = \left[ {\zeta}_{1}^\mathrm{T}, \ldots, {\zeta}_{N}^\mathrm{T} \right]^\mathrm{T}$, and the mapping $U$ is selected such that 
		\begin{align*}
			I_{N}\otimes S & +\Lambda\mathcal{H}\otimes I_{q} \notag \\
			& = \bigg( I_{N}\otimes \Big[ S-(\lambda_{M}(S)+r) I_{q} \Big] \bigg) (U\otimes I_{q}). 
		\end{align*}
		$I_{N}\otimes S  +\Lambda\mathcal{H}\otimes I_{q} = \left( I_{N}\otimes [ S-(\lambda_{M}(S)+r) I_{q} ] \right) (U\otimes I_{q})$. 
		Note that using Assumption 5, the graph $\mathcal{G}$ contains no loop, so the nodes can be labeled as $i >j$ if $(v_i, v_j) \in \mathcal{E}$ and $\mathcal{H}$ is upper triangular with diagonal element $d_i >0$, which ensures such a matrix $U$ exists.

		Next, using \eqref{eq:9}, it follows that
		\begin{align}
			\dot{\varsigma}(t)& =\left(I_{N}\otimes S+\Lambda\mathcal{H}\otimes I_{q}\right)\varsigma (t)\notag \\
			& = \Big( I_{N}\otimes [ S-(\lambda_{M}(S)+r) I_{q} ] \Big) (U\otimes I_{q}) \varsigma(t) \notag \\
			& =  I_{N}\otimes [ S-(\lambda_{M}(S)+r) I_{q} ] {\zeta}(t), \quad t \ge 0, 
		\end{align}
		which implies that $\dot{\varsigma}_{i}(t)  = \left[ S-(\lambda_{M}(S)+r)I_{q} \right] {\zeta}_{i}(t)$, $t \ge 0$ for $i= 1, \ldots, N$. 
		
		Now, we design ${\zeta}_{i0}$ for \eqref{eq:10} as
		\begin{align} \label{eq:17}
			{\zeta}_{i0} & = {\zeta}_{j0}, \quad i, j =1, \ldots, N. 
		\end{align}
		which indicates that ${\zeta} (t) = (\mathbf{1}_{N} \otimes I_q ){\zeta}_{i}(t)$, $i = 1, \ldots, N$. Further, 
		$\varsigma_{i}(t)  = (T_{i} \otimes I_q )  \varsigma(t) = (T_i U^{-1} \mathbf{1}_{N} \otimes I_q) {\zeta}_{i}(t)$,  $t \ge 0$, 
		and 
		\begin{align*}
			\varsigma_{i}(t) & = (T_{i} \otimes I_q )  \varsigma(t) =(T_{i} \otimes I_q ) (U^{-1}\otimes I_{q}) \zeta(t) \notag \\
			& = (T_i U^{-1} \mathbf{1}_{N} \otimes I_q) {\zeta}_{i}(t), \quad t \ge 0, 
		\end{align*}
		where $T_i \in \mathbb{R}^{1\times N}$ denotes the vector $[0 \cdots 0~1~0 \cdots 0]$ with $i$th component $1$ and the other components $0$. 
		
		Then for \eqref{eq:6}, we have 
		\begin{align}
			\sum_{j=0}^{N} \rho_{ij}(\xi_{i}-\xi_{j}) =T_{i} (\mathcal{H}U^{-1}\mathbf{1}_{N})\zeta_i.  \label{eq:204} 
		\end{align}
		and 
		\begin{align}
			\sum_{j=0}^{N} \rho_{ij}(\xi_{i}-\xi_{j}) 
			& = \rho_{i0} {\varsigma}_{i} + \sum_{j=1}^{N} \rho_{ij} ( {\varsigma}_{i}- {\varsigma}_{j} ) \notag  \\
			& =T_{i}(\mathcal{A}_{0}U^{-1}\mathbf{1}_{N})\zeta_{i} +  (\mathcal{L}_{s}\otimes I_{q}) \varsigma  \notag \\
			& =T_{i}(\mathcal{A}_{0}U^{-1}\mathbf{1}_{N})\zeta_{i} +  (\mathcal{L}_{s}\otimes I_{q})(U^{-1}\otimes I_{q})\zeta  \notag \\
			& =T_{i}(\mathcal{A}_{0}U^{-1}\mathbf{1}_{N}) \zeta_{i}+ T_i (\mathcal{L}_{s}U^{-1} \mathbf{1}_{N}) \zeta_i  \notag \\
			& =T_{i} (\mathcal{H}U^{-1}\mathbf{1}_{N})\zeta_i.  \label{eq:24} 
		\end{align}

		Now, we define
		\begin{align*}
			\tilde{x}_{i}(t) & \buildrel \vartriangle \over = x_{i}(t) - \Pi_{i}\xi_{i}(t),   \\
			\tilde{u}_{i}(t) & \buildrel \vartriangle \over = u_{i}(t) - \Gamma_{i}\xi_{i}(t), 
		\end{align*}
		and rewrite \eqref{eq:1} as $\dot{x}_{i}(t)=A_{i}x_{i}(t)+B_{i}u_{i}(t)+E_{i}\xi_{i}(t)-E_{i}\varsigma{}_{i}(t)$, $t \ge 0$. 
		Then using \eqref{eq:6} and \eqref{eq:24}, it follows that
		\begin{align}  \label{eq:25}
			\dot{\tilde{x}}_{i}(t) 
			& =  A_{i}x_{i}(t)+B_{i}u_{i}(t)+E_{i}\xi_{i}(t)-E_{i}\varsigma_{i}(t)-\Pi_{i}\dot{\xi}_{i}(t) \notag \\
			= &A_{i}\tilde{x}_{i}+A_{i}\Pi_{i}\xi_{i}+B_{i}\tilde{u}_{i}+B_{i}\Gamma_{i}\xi_{i}+E_{i}\xi_{i}-E_{i}\varsigma_{i}-\Pi_{i}\dot{\xi}_{i}  \notag \\
			& = A_{i}\tilde{x}_{i}(t)+B_{i}\tilde{u}_{i}(t)-E_{i}\varsigma_{i}(t)+(A_{i}\Pi_{i}+B_{i}\Gamma_{i}+E_{i})\xi_{i}(t)  \notag \\
			& \quad -\Pi_{i}\Bigg[ S\xi_{i}(t)+\alpha_{i}\bigg(\sum_{j=0}^{N}\rho_{ij}(\xi_{i}(t)-\xi_{j}(t))\bigg) \Bigg] \notag \\
			& =  A_{i}\tilde{x}_{i}(t) + B_{i}\tilde{u}_{i}(t) - E_{i}\varsigma_{i}(t) \notag \\
			& \quad - \alpha_{i}\Pi_{i}\bigg[ \sum_{j=0}^{N}\rho_{ij}(\xi_{i}(t) - \xi_{j}(t)) \bigg] \notag \\
			&  = A_{i} \tilde{x}_{i}(t) + B_{i}\tilde{u}_{i}(t) -\varPhi_{i} {\zeta_{i}}(t), 
		\end{align}
		where $\varPhi_{i} \buildrel \vartriangle \over = E_{i}(T_{i}U^{-1}\mathbf{1}_{N})+\alpha_{i}\Pi_{i} T_{i} (\mathcal{H}U^{-1}\mathbf{1}_{N})$.
		
		For the output error in \eqref{eq:4}, note that $e_{i}(t)=C_{i}x_{i}(t)+D_{i}u_{i}(t)-F_{i}\xi_{i}(t)+F_{i}\varsigma_{i}(t)$, $t \ge 0$, and then, using \eqref{eq:22-2}, it follows that 
		\begin{align}  \label{eq:27} 
			e_{i} (t) 
			&  =C_{i}\tilde{x}_{i}(t)+D_{i}\tilde{u}_{i}(t)+(C_{i}\Pi_{i}+D_{i}\Gamma_{i}-F_{i})\xi_{i}(t)+F_{i}\varsigma_{i}(t) \notag \\
			& =C_{i}\tilde{x}_{i}(t) +D_{i}\tilde{u}_{i}(t) +F_{i}\varsigma_{i}(t) \notag \\
			& = C_{i}\tilde{x}_{i} (t) +D_{i}\tilde{u}_{i}(t) -\varPsi_{i}{\zeta_{i}} (t), 
		\end{align}
		where $\varPsi_{i} \buildrel \vartriangle \over = -F_{i}(T_{i}U^{-1} \mathbf{1}_{N})$. 
		
		For the control law $\tilde{u}_{i}(t)$, $t, \ge 0$, we use \eqref{eq:21-3} and rewrite it as 
		\begin{align}  \label{eq:28}
			\tilde{u}_{i}(t) & =u_{i}(t)- \Gamma_{i}\xi_{i}(t) \notag \\
			& =-K_{i,1}\tilde{x}_{i}(t) - K_{i,3}\zeta_{i}(t). 
		\end{align}
		
		Next, defining ${X_{i}}(t) \buildrel \vartriangle \over = \left[\zeta_{i}^\mathrm{T}(t), \tilde{x}_{i}^\mathrm{T}(t)\right] ^\mathrm{T}$, $t \ge 0$, and using \eqref{eq:10}, \eqref{eq:25}, and \eqref{eq:27},  it follows that 
		\begin{align}
			\dot{{X}}_{i} (t) & = {A}_{ic}  {X}_{i} (t) + {B}_{ic} \tilde{u}(t), \quad t \ge 0,   \label{eq:34-1}  \\
			e_{i}(t) & ={C}_{ic}{X}_{i}(t) + {D}_{ic} \tilde{u}(t),  \label{eq:34-2} 
		\end{align}
		where ${X_{i0}} = {X_{i}}(0) = \left[\zeta_{i0}^\mathrm{T}, \tilde{x}_{i}^\mathrm{T}(0)\right] ^\mathrm{T}$ and 
		\begin{align*}
			{A}_{ic} & = \begin{bmatrix}
				S-(\lambda_{M}(S)+r)I_{q}  & 0 \\
				- \varPhi_{i} & A_{i} \end{bmatrix}, \quad B_{ic} = \begin{bmatrix} 0 \\ B_i \end{bmatrix}, \\
			C_{ic} & =\begin{bmatrix} -\varPsi_{i} & C_{i} \end{bmatrix}, \quad D_{ic} = D_i.
		\end{align*}
		Substituting \eqref{eq:28} into \eqref{eq:34-1} and \eqref{eq:34-2} yields the following closed-loop dynamics
		\begin{align}
			\dot{{X}}_{i} (t) & = \big( {A}_{ic} - {B}_{ic} {K}_{ic}  \big) {X}_{i} (t), \quad t \ge 0, \label{eq:34}  \\
			e_{i}(t) & = \big( {C}_{ic} - {D}_{ic} {K}_{ic}  \big) {X}_{i} (t) \label{eq:35} 
		\end{align}
		where ${K}_{ic} \buildrel \vartriangle \over = \begin{bmatrix} K_{i,3} & K_{i,1} \end{bmatrix} \in \mathbb{R}^{m_i \times (q + n_i)}$. 
		
		Notice that for each $i = 1, \ldots, N$, the matrix 
		\begin{align*}
			{A}_{ic} - {B}_{ic} {K}_{ic}  = \begin{bmatrix}
				S-(\lambda_{M}(S)+r)I_{q}  & 0 \\
				- \varPhi_{i}  - B_i K_{i,3}  & A_{i} - B_i K_{i,1} \end{bmatrix}, 
		\end{align*}
		is Hurwitz stable. Then using similar arguments as in \cite{Huang2004}, the augmented system \eqref{eq:34} is exponentially stable. Thus, the state transition matrix $\Phi_{i}(t, \tau)$ of \eqref{eq:34} satisfies $|\Phi_{i}(t, \tau)|\le k_{i}e^{-\gamma_{i}(t - \tau)}$, $ t\ge \tau > 0$, for $k_{i},\gamma_{i}\in\mathbb{R}_+$. 
		Next, using \eqref{eq:35} with $\bar{C}_{ic} \buildrel \vartriangle \over = {C}_{ic} - {D}_{ic} {K}_{ic}$ yields $|e_{i}(t)|\le |\bar{C}_{ic})| |X_{i}(\tau)| k_{i}e^{-\gamma_{i}(t-\tau)} \to 0$ as $t \to \infty$. Finally, the tracking errors satisfy $\mathrm{lim}_{t\to\infty}e_{i}(t)=0$ for $i = 1, \ldots, N$, which complete the proof. 
	
	Theorem~1 indicates that 
	the MASs given by \eqref{eq:1} and \eqref{eq:2} with \eqref{eq:6},  \eqref{eq:10}, and \eqref{eq:11}, not only can achieve the asymptotic stability; 
	but also can achieves the output synchronization \cite{Huang2004} over the communication network $\mathcal{G}$.
	We note that there is not a unique solution of $K_{i,1}$, $K_{i,2}$, and $K_{i,3}$ in  \eqref{eq:11} that satisfies conditions of Theorem 1.

	\section{Optimal Synchronization Control Via distributed PI Technique}  \label{sec4}
	In this section, we present the optimal synchronization control via optimizing the feedback gains given in \eqref{eq:11} and minimizing the output tracking errors. Then we obtain the distributed optimal control policy via online PI technique \cite{Sutton1998,zhou2020secure}, which provides a unique and optimal learning solution for each agent $i$.  
	
	To make the control protocols \eqref{eq:6}, \eqref{eq:10}, and \eqref{eq:11} be unique with optimality and ameliorate the closed-loop performance, we introduce a cost function associated with \eqref{eq:4} as
	\begin{align}  \label{eq:38}
		J_{i} \buildrel \vartriangle \over  =J_{i}(X_{i0}, \tilde u_{i}(\cdot)) =\int_{0}^{\infty}e_{i}^\mathrm{T} (t) e_{i}(t) \mathrm{d}t.
	\end{align}
	Then using Theorem 1 and \eqref{eq:34-1}, a dynamical feedback controller is designed with the form
	\begin{align} \label{eq:36}
		\tilde{u}_i (t)= - K_{ic} X_i(t),  \quad t \ge 0. 
	\end{align}
	In this case, using \eqref{eq:34-2}, \eqref{eq:38} can be rewritten as
	\begin{align} \label{eq:37}
		J_{i} &  =\int_{0}^{\infty}(C_{ic}X_{i}(t)+D_{ic}\tilde{u}_{i}(t))^\mathrm{T}(C_{ic}X_{i}(t)+D_{ic}\tilde{u}_{i}(t)) \mathrm{d}t  \notag \\
		& =\int_{0}^{\infty}X_{i}^\mathrm{T}(t) (C_{ic} - D_{ic}K_{ic})^\mathrm{T}(C_{ic} - D_{ic}K_{ic})X_{i}(t) \mathrm{d}t. 
	\end{align}
	
	For the cost \eqref{eq:37}, we note that it is a quadratic function with cross-product term. For the $i$th subsystem given by \eqref{eq:34-1} and \eqref{eq:34-2}, the control policy given by \eqref{eq:11} with $K_{ic}$ determined by Theorem 1 is not optimal, which does not consider minimization of the cost $J_{i}$ in \eqref{eq:38}.    
	Next, given the system \eqref{eq:34-1}-\eqref{eq:34-2} with \eqref{eq:38}, the Hamiltonian function in terms of the control policy $\tilde u_{i}$ is defined as
	\begin{align}  \label{eq:39}
		\mathcal{H}_{i}(X_{i},\tilde{u}_{i}) & =e_{i}^\mathrm{T}e_{i}+X_{i}^\mathrm{T}(A_{ic}^\mathrm{T}P_{ic} +P_{ic}A_{ic})X_{i} \notag \\
		& \quad +2X_{i}^\mathrm{T}P_{ic}B_{ic}\tilde{u}_{i}
	\end{align}
	where $P_{ic} = P_{ic}^\mathrm{T} \succ 0$. 
	
	Now, using the the optimal control theory and stationary condition for optimality $\nabla_{\tilde{u}_{i}}\mathcal{H}_{i}(X_{i},\tilde{u}_{i})=0$, where $\nabla_{\tilde{u}}(*)$ is the Fr\'{e}chet derivative of the function $*$ at $\tilde{u}$, the optimal controller is obtained with the form
	\begin{align}  \label{eq:40}
		\tilde{u}_{i}^{\star} (t) =-K_{ic}^{\star}X_{i}(t),
	\end{align}
	where 
	\begin{align}   \label{eq:41}
		K_{ic}^{\star}=(D_{ic}^\mathrm{T}D_{ic})^{-1}(D_{ic}^\mathrm{T}C_{ic}+B_{ic}^\mathrm{T}P_{ic}^\star). 
	\end{align}
	and the associated optimal cost matrix $P_{ic}^\star = P_{ic}^{\star\mathrm{T}} \succ 0$  satisfies 
	\begin{align}   \label{eq:43}
		0= & A_{ic}^\mathrm{T}P_{ic}^\star +P_{ic}^\star A_{ic}+C_{ic}^\mathrm{T}C_{ic} -(D_{ic}^\mathrm{T}C_{ic}+B_{ic}^\mathrm{T}P_{ic}^\star)^\mathrm{T}  \notag \\
		& \times (D_{ic}^\mathrm{T}D_{ic})^{-1}  (D_{ic}^\mathrm{T}C_{ic}+B_{ic}^\mathrm{T}P_{ic}^\star). 
	\end{align}
	
	Next, combining \eqref{eq:41} and \eqref{eq:43} yields the following algebraic Riccati equation (ARE)
	\begin{align}   \label{eq:42}
		0 =  & \left( A_{ic}-B_{ic}K_{ic}^{\star} \right)^\mathrm{T}P_{ic}^\star+P_{ic}^\star (A_{ic}-B_{ic}K_{ic}^\star)   \notag \\
		& +(C_{ic}-D_{ic}K_{ic}^\star)^\mathrm{T}(C_{ic}-D_{ic} K_{ic}^\star). 
	\end{align}
	where $P_{ic}^\star = P_{ic}^{\star\mathrm{T}} \succ 0$ and $K_{ic}^\star$ denotes the unique solutions.

	Note that \eqref{eq:43} is nonlinear in $P_{ic}^\star$, 
	we employ the PI technique \cite{Sutton1998} to approximate $P_{ic}^\star$ by solving linear Lyapunov equations iteratively. 
	In this case, we can learn the optimal feedback gains instead of directly solving the ARE \eqref{eq:43}. Finally, we obtain the PI-based distributed optimal control algorithm summarized in Algorithm 1. 
	


	\medskip
	\hrule \vspace{0.15cm}
	$\textbf{Algorithm 1.}$ Adaptive learning the optimal control protocol \eqref{eq:11} via PI technique. 
	\hrule \vspace{0.15cm}
	\begin{enumerate}
		\item Select a constant $r >0$ and a sufficiently small constant $\epsilon>0$. Let $i \leftarrow1$ 
		\item $\textbf{Repeat:}$
		\begin{enumerate}
			\item Find a pair $(\Pi_{i},\Gamma_{i})$ to satisfy \eqref{eq:22-1} and \eqref{eq:22-2} and an appropriate parameter $\alpha_{i}$ to satisfy \eqref{eq:21-1}. 
			\item Find a stabilizing $K_{ic}^{[0]}$ according to Theorem 1. Let $k\leftarrow0$. 
			\item $\textbf{Repeat:}$
			\begin{enumerate}
				\item Solve $P_{ic}^{[k]}$ from
				\begin{align}
					0 =  & \Big( \bar{A}_{ic}^{[k]} \Big)^\mathrm{T}P_{ic}^{[k]}+P_{ic}^{[k]} \bar{A}_{ic}^{[k]}   \notag \\
					&  + \Big(C_{ic} - D_{ic}K_{ic}^{[k]} \Big)^\mathrm{T} \Big(C_{ic}  - D_{ic} K_{ic}^{[k]} \Big)  \label{eq:44}
				\end{align}
				where 
				\begin{align}  \label{eq:52}
					\bar{A}_{ic}^{[k]} \buildrel \vartriangle \over = A_{ic} - B_{ic}K_{ic}^{[k]}. 
				\end{align}
				\item Solve $K_{ic}^{[k+1]}$ from
				\begin{align}
					K_{ic}^{[k+1]}= & - \left( D_{ic}^\mathrm{T}D_{ic} \right)^{-1} \Big( D_{ic}^\mathrm{T}C_{ic}+B_{ic}^\mathrm{T}P_{ic}^{[k]} \Big)  \label{eq:45}
				\end{align}
				\item Let $k\leftarrow k+1$. 
			\end{enumerate}
			\item $\textbf{until}$ $\big| K_{ic}^{[k+1]}-K_{ic}^{[k]} \big|<\epsilon$.
		\end{enumerate}
		\item Let $i\leftarrow i+1$. 
		\item $\textbf{until}$ $i=N$
	\end{enumerate}
	\hrule \vspace{0.15cm}
	
	\medskip

	\textbf{Theorem~2. }   
	Consider the disturbed MASs given by (1) and (2) and choose $K_{ic}^{[0]}$ by using Theorem 1. Then for each iteration $k = 0, 1, \ldots$ of Algorithm 1, we have that 
	\begin{enumerate}
		\item $\bar{A}_{ic}^{[k]}$ is Hurwitz stable.
		\item $P_{ic}^\star \le P_{ic}^{[k+1]} \le P_{ic}^{[k]} \le \cdots \le P_{ic}^{[0]}$.
		\item $\mathrm{lim}_{k\to\infty}P_{ic}^{[k]}=P_{ic}^\star$,  $\mathrm{lim}_{k\to\infty}K_{ic}^{[k]}=K_{ic}^{^\star}$.  
	\end{enumerate}  

	\emph{Proof. }
		The proof follows by mathematical induction. Specifically: 
		\begin{enumerate}
			\item For $k = 0$, using Theorem 1, we have that $K_{ic}^{[0]}$ is chosen such that the matrix $\bar{A}_{ic}^{[0]}$ is Hurwitz stable. 
			\item Suppose that for some  $k >0$, $\bar{A}_{ic}^{[k]}$ is Hurwitz stable and $P_{ic}^{[k]} \le \cdots \le P_{ic}^{[0]}$. This indicates that $\bar{A}_{ic}^{[k]}$ has eigenvalues with negative real parts. 
		\end{enumerate}
		Next, we consider the case with $k+1$. 
		Using \eqref{eq:44} yields the associated cost matrix $P_{ic}^{[k+1]}$ and using similar arguments as in \cite{Kleinman1968}, we obtain that $P_{ic}^{[k+1]} \le P_{ic}^{[k]}$. Since  $P_{ic}^{[k+1]}$ is bounded, it has finite norm. So $P_{ic}^{[k+1]}$ satisfy \eqref{eq:44} with $k+1$ and $K_{ic}^{[k+1]}$ is uniquely determined by 
		\eqref{eq:45}. Using monotonic convergence of matrices series $\{ P_{ic}^{[k]}  \}_{k=1}^{\infty}$, we have that $\mathrm{lim}_{k\to\infty}P_{ic}^{[k]}=P_{ic}^\star$. Since for each $k = 0, 1, \ldots$, $K_{ic}^{[k+1]}$ is the unique positive define solution of \eqref{eq:45}, we obtain that  $\mathrm{lim}_{k\to\infty}K_{ic}^{[k]}=K_{ic}^{^\star}$. 
		
		Finally, the proof is completed.

	In~Algorithm 1, we develop an optimal synchronization control protocol by minimizing the output tracking errors and using distributed PI technique  \cite{Sutton1998}. 
	Although Algorithm 1 requires the feedback gains $K_{i,1}$, $K_{i,2}$, and $K_{i,3}$ for an initially stabilizing control policy $K_{ic}^{[0]}$, Theorem 1 provides the basic conditions for designing such a policy. Note that it is not the unique optimal solution. Algorithm 1 uses the iterative equations \eqref{eq:44} and \eqref{eq:45} to obtain the $K_{ic}^{^\star}$, which is an online adaptive learning solution. This provides a unique learning solution to our \emph{Problem 1}.

	%
	%

	\section{Simulation Example} \label{sec5}
\begin{figure}[bt]
		\centering
		\includegraphics[width=2.0in] {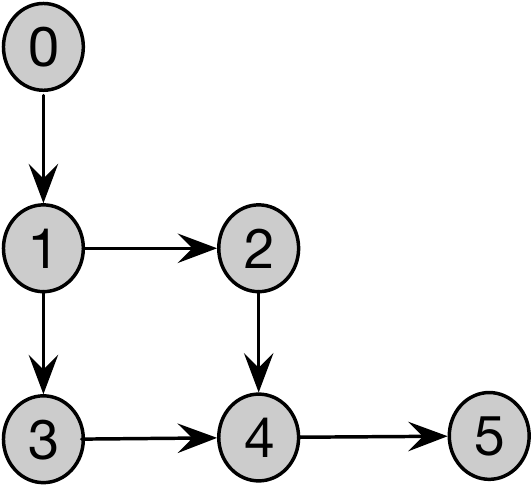}\\
		\caption{The communication network with the graph $\mathcal{G}$.} 
		\label{fig:1}
	\end{figure}

	To illustrate the key ideas presented in this paper, we provide an illustrative numerical example representing  a heterogeneous multi-agent system consisting of six agents. The communication network with the graph $\mathcal{G}$ is given in Figure~\ref{fig:1}.
	The system dynamics are given in the form of \eqref{eq:1}, \eqref{eq:3}, and \eqref{eq:4} with 
	\begin{description}
		\item[\emph{Agent 1}: ] 
		\begin{align*}
			A_1  & =  \begin{bmatrix} -1 & 0 & 0.5 \\0 & -1  & 0 \\0 & 0 & -1 \end{bmatrix} ,  \quad 
			B_1 = \begin{bmatrix} 0 & 0 \\ 0 & 1.5 \\ 1 &0 \end{bmatrix},   \quad
			E_1 = \begin{bmatrix} 1 & 0 \\ 0 & 0.5 \\ 1 & 0 \end{bmatrix},\\  
			C_1  & =  \begin{bmatrix} 1 & 0 & 0 \\ 0 & 2 & 0 \end{bmatrix}, \quad
			D_1  =  \begin{bmatrix} 0.5 & 0 \\ 0.5 & 1.5 \end{bmatrix}, \quad
			F_1 = \begin{bmatrix} 0.5 & 0 \\ 0 & 0.5 \end{bmatrix}.
		\end{align*}
		\item[\emph{Agent 2}: ] 
		\begin{align*}
			A_2 & =  \begin{bmatrix} -1 & 0 & 1 \\0 & -1 &0 \\0 & 0 & -1 \end{bmatrix}, \quad
			B_2 = \begin{bmatrix} 0 & 0\\ 0 & 2 \\ 1 & 0 \end{bmatrix} , \quad
			E_2 =  \begin{bmatrix} 1 & 0 \\ 0 & 1 \\ 1 & 0 \end{bmatrix}, \\  
			C_2 & =   \begin{bmatrix} 1.5 & 0 & 0 \\0 &2 &1 \end{bmatrix}, \quad
			D_2 = \begin{bmatrix} 1 & 0 \\ 1 & 2 \end{bmatrix}, \quad
			F_2 =\begin{bmatrix} 1 & 0 \\ 0 &3 \end{bmatrix}.
		\end{align*}
		\item[\emph{Agent 3}: ] 
		\begin{align*}
			A_3  & =  \begin{bmatrix} -1 & 0 & 1.5 \\0 & -1 &0 \\0 & 0 & -1 \end{bmatrix},   \quad
			B_3 = \begin{bmatrix} 0 & 0 \\ 0 & 4.5 \\ 1 &0 \end{bmatrix},   \quad
			E_3 = \begin{bmatrix} 1 & 0 \\ 0 & 1.5 \\ 1& 0 \end{bmatrix}, \\  
			C_3 & =  \begin{bmatrix} 1.5 & 0 & 0 \\ 0 & 2.5 & 0 \end{bmatrix},  \quad
			D_3 =  \begin{bmatrix} 1.5 & 0 \\ 0.5 & 2 \end{bmatrix},   \quad
			F_3 = \begin{bmatrix} 1.5 & 0 \\ 0 & 2 \end{bmatrix}.
		\end{align*}
		\item[\emph{Agent 4}: ] 
		\begin{align*}
			A_4 & =  \begin{bmatrix} -1 &  0 & 2 \\0 & -1 & 0 \\ 0 & 0 &-1 \end{bmatrix},   \quad
			B_4 = \begin{bmatrix} 0 & 0 \\ 0 &1 \\1 &0 \end{bmatrix} ,   \quad
			E_4 = \begin{bmatrix} 1 & 0 \\ 0 & 2 \\ 1 & 0 \end{bmatrix}, \\
			C_4 & =  \begin{bmatrix} 2 & 0 &0 \\ 0 & 2.5 & 0 \end{bmatrix},   \quad
			D_4 =  \begin{bmatrix} 2 & 0 \\ 0.5 & 2 \end{bmatrix},    \quad
			F_4 = \begin{bmatrix} 2 & 0 \\ 0 & 1 \end{bmatrix}.
		\end{align*}
		\item[\emph{Agent 5}: ] 
		\begin{align*}
			A_5 & =  \begin{bmatrix} -1 & 0 & 2.5 \\0 & -1 & 0 \\ 0 & 0 & -1 \end{bmatrix},   \quad
			B_5 = \begin{bmatrix} 0 & 0 \\0 & 2.5\\ 1 &0 \end{bmatrix} , \quad
			E_5 = \begin{bmatrix} 1 & 0 \\0 & 2.5\\ 1 &0 \end{bmatrix} ,  \\
			C_5 & =  \begin{bmatrix} 2.5 & 0 & 0 \\ 0 & 3 &0  \end{bmatrix}, \quad
			D_5 =  \begin{bmatrix} 2.5 & 0 \\0.5 & 2.5 \end{bmatrix}, \quad
			F_5 = \begin{bmatrix} 2.5 & 0 \\ 0 & 2.5 \end{bmatrix}.
		\end{align*}
	\end{description}
	and the leading agent is described in the form of \eqref{eq:2} with
	\begin{align*}
		S =  \begin{bmatrix} 1 & 0 \\ 0 & 1 \end{bmatrix}. 
	\end{align*}
	The Laplacian matrix of the communication graph $\mathcal{G}$ in Figure~\ref{fig:1} is given by
	\begin{align*}
		\mathcal{L} = \begin{bmatrix} 0 & 0 & 0 & 0 & 0 & 0 \\ -1 & 1 & 0 & 0 & 0 & 0 \\ 0 & -1 & 1 & 0 & 0 & 0 \\
			0 & -1 & 0 & 1 & 0 & 0 \\ 0 & 0 & -1 & -1 & 2 & 0 \\ 0 & 0 & 0 & 0 &  -1& 1 \end{bmatrix}. 
	\end{align*}
	It can be verified that Assumptions 1-5 hold. The solutions of the output regulator equations \eqref{eq:22-1} and  \eqref{eq:22-2} for the five agents are given by
	\begin{align*}
		\Pi_1 & = \begin{bmatrix} 0.6 & 0 \\ 0.025 & 0.25 \\ 0.4 & 0 \end{bmatrix}, 
		\quad \Gamma_1 = \begin{bmatrix} -0.2 & 0 \\ 0.0333 & 0 \end{bmatrix} , \\
		\Pi_2 & = \begin{bmatrix} 0.7273 & 0 \\ -0.0909 & 1 \\ 0.4545 & 0 \end{bmatrix}, 
		\quad \Gamma_2 = \begin{bmatrix} -0.0909 & 0 \\ -0.0909 & 0.5 \end{bmatrix} , \\
		\Pi_3 & = \begin{bmatrix} 0.9091 & 0 \\ -0.0134 & 0.7869 \\ 0.5455 & 0 \end{bmatrix}, 
		\quad \Gamma_3 = \begin{bmatrix} 0.0909 & 0 \\ -0.006 & 0.0164 \end{bmatrix} , \\
		\Pi_4 & = \begin{bmatrix} 1 & 0 \\ 0 & 0.7692 \\ 0.5 & 0 \end{bmatrix}, 
		\quad \Gamma_4= \begin{bmatrix} 0 & 0 \\ 0 & -0.4615 \end{bmatrix} , \\
		\Pi_5 & = \begin{bmatrix} 1.0769 & 0 \\ 0.0077 & 1 \\ 0.4615 & 0 \end{bmatrix}, 
		\quad \Gamma_5 = \begin{bmatrix} - 0.0769 & 0 \\ 0.0062 & -0.2 \end{bmatrix}. 
	\end{align*}
	
\begin{figure}[bt]
		\centering
		\includegraphics[width=10cm]{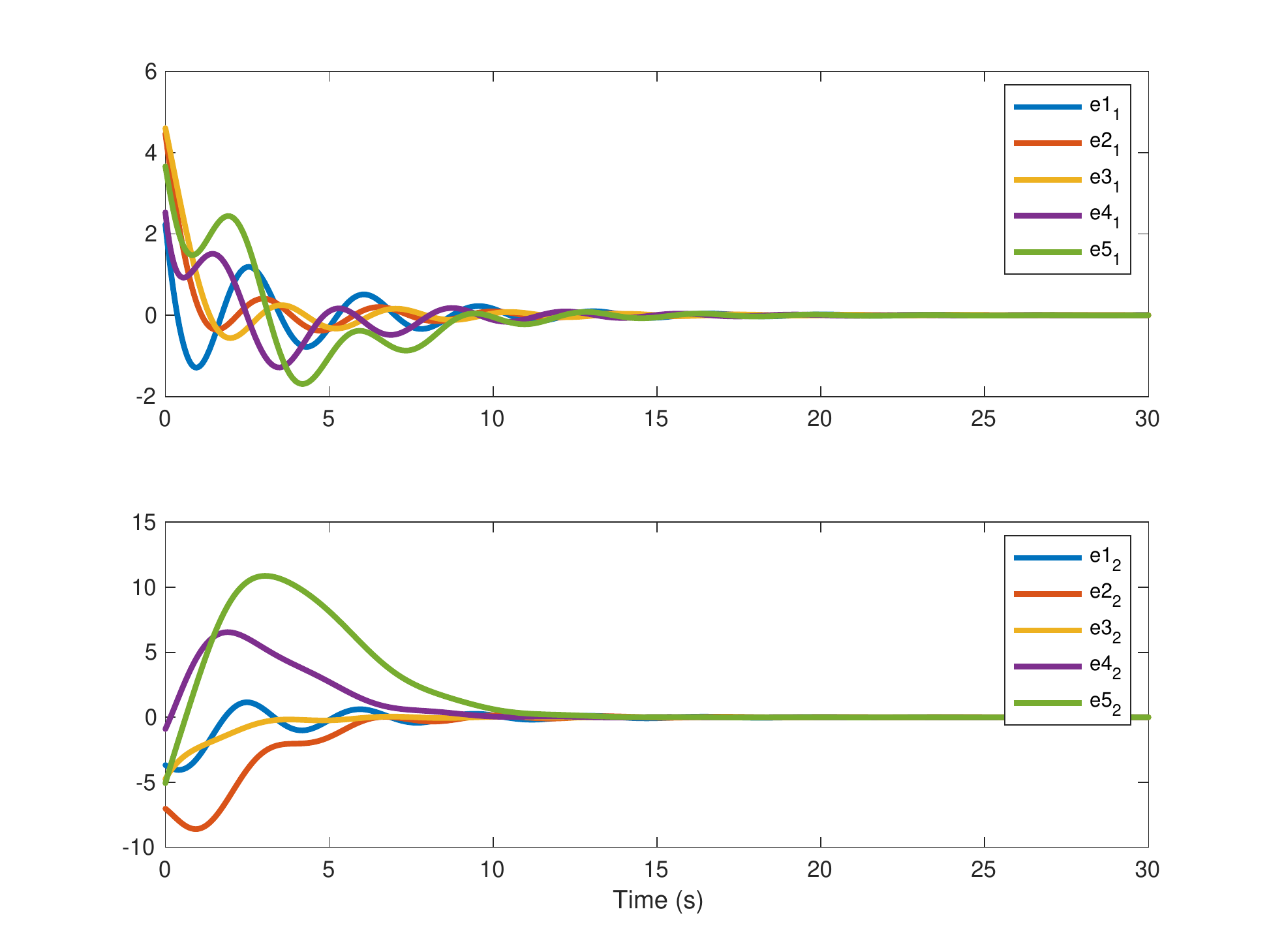}\\
		\caption{Tracking errors of the five vehicles under initially distributed control  protocols. } 
		\label{fig:2}
	\end{figure}

	Now, according to Theorem 1, we let $r =1$ and design an initially distributed control protocol by \eqref{eq:11} with the gain matrices given by
	\begin{align*}
		K_{1,1} & = \begin{bmatrix} 4 & 0 & 3 \\ 0 & 0 & 0 \end{bmatrix}, 
		\quad K_{1, 2} = \begin{bmatrix} -3.4 & 0 \\ -0.0333 & 0 \end{bmatrix} , \\
		K_{2,1} & = \begin{bmatrix} 2 & 0 & 3 \\ 0 & 0 & 0 \end{bmatrix}, 
		\quad K_{2, 2} = \begin{bmatrix} -2.7273 & 0 \\ 0.09 & -0.5 \end{bmatrix} , \\
		K_{3,1} & = \begin{bmatrix} 1.3333 & 0 & 3 \\ 0 & 0 & 0 \end{bmatrix}, 
		\quad K_{3, 2} = \begin{bmatrix} -2.9394 & 0 \\ 0.006 & -0.0164 \end{bmatrix} , \\
		K_{4,1} & = \begin{bmatrix} 1 & 0 & 3 \\ 0 & 0 & 0 \end{bmatrix}, 
		\quad K_{4, 2} = \begin{bmatrix} -2.5 & 0 \\ 0 & 0.4615 \end{bmatrix} , \\
		K_{5,1} & = \begin{bmatrix} 0.8 & 0 & 3 \\ 0 & 0 & 0 \end{bmatrix}, 
		\quad K_{5, 2} = \begin{bmatrix} -2.1692 & 0 \\ -0.0062 & 0.2 \end{bmatrix}, \\
		K_{1,3} & =K_{2,3} = K_{3,3} = K_{4,3} = K_{5,3} = 0_{2 \times 2}. 
	\end{align*}
	The initial values for $w_0$, $x_{i0}$, and $\xi_{i0}$ are randomly selected around the origin with $w_0 \neq 0$. In this case, the tracking  performance under the initially distributed control protocols is shown in Figure~\ref{fig:2}.
	
	Next, using Algorithm 1, we obtain the optimal solutions under the initial control protocols with ${K}_{ic}^{[0]} =\left[K_{i,3}, K_{i,1} \right]$ for each $i = 1, \ldots, 5$. In this case, the optimal feedback control gains $K_{ic}^\star$ are obtained as
	\begin{align*}
		K_{1c}^\star & = \begin{bmatrix} 0.369 & 0.0388 & 0.7187 & 0.1552 & 0.0404 
			\\ -0.1069 & 0.2801 & -0.2089 & 1.1204 & -0.0135 \end{bmatrix}, \\
		K_{2c}^\star & = \begin{bmatrix} 0.5255 & 0.7417 & 0.2852 & 0.1660 & 0.1382 
			\\ -0.1873 & 2.3818 & -0.0855 & 0.5209 & 0.2368 \end{bmatrix}, \\
		K_{3c}^\star & = \begin{bmatrix} 1.0685 & 0.1592 & 0.3587 & 0.0628 & 0.0762 
			\\ -0.1668 & 1.7802 & -0.0551 & 0.7645 & -0.0159 \end{bmatrix}, \\
		K_{4c}^\star & = \begin{bmatrix} 2.2503  & 0.1967& 0.3442 & 0.0475 & 0.1169
			\\-0.1979 & 0.9424 &-0.0313 & 0.4658 &-0.0161 \end{bmatrix}, \\
		K_{5c}^\star & = \begin{bmatrix}  1.0057 & 0.1640 & 0.0875 & 0.0116  & 0.0697
			\\-0.0147 & 1.1484 &-0.0021 & 0.1485 &-0.0028 \end{bmatrix}. 
	\end{align*}

\begin{figure}[bt]
		\centering
		\includegraphics[width=10cm] {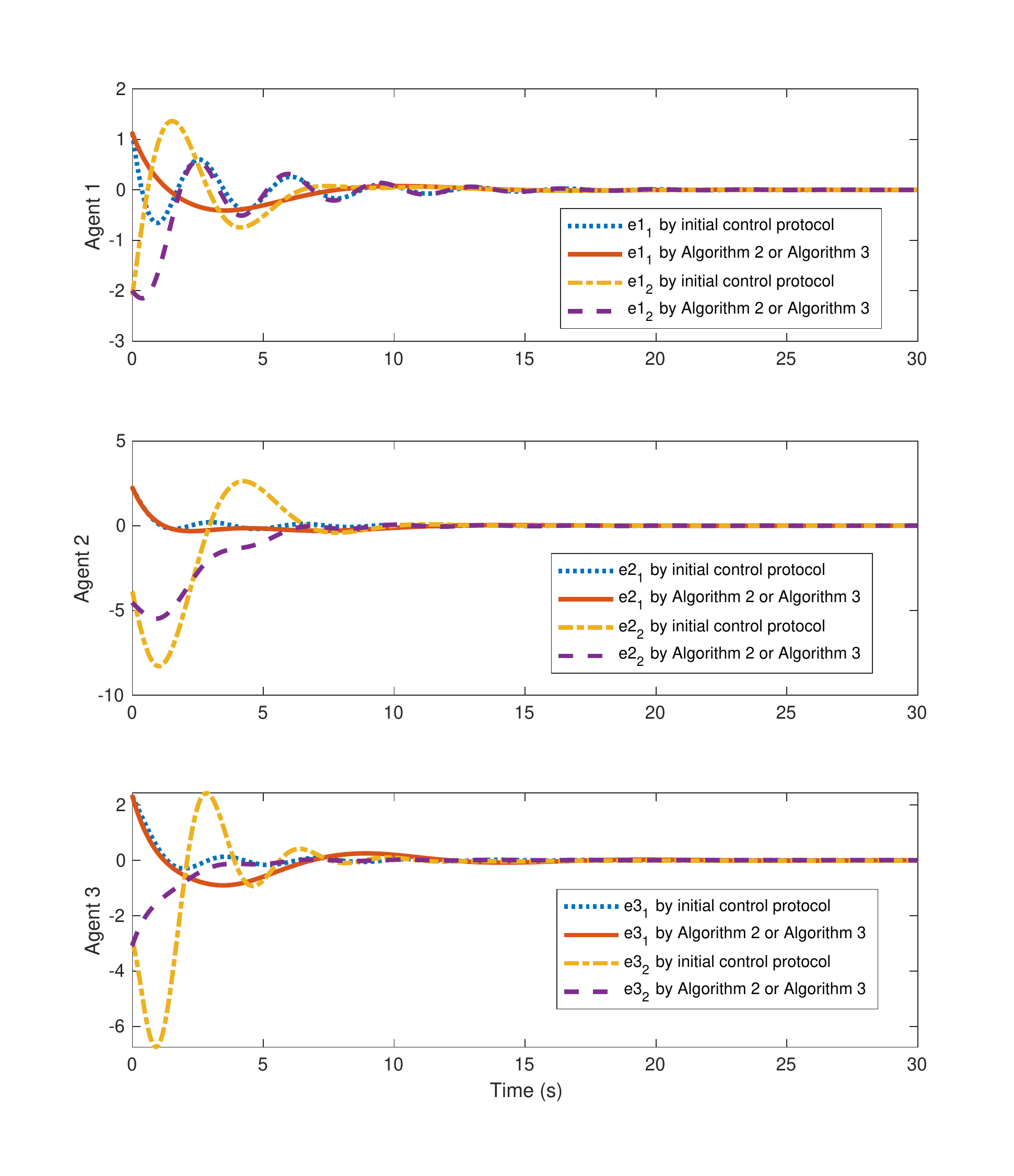}\\
		\caption{Comparisons of system performance under the initial and optimal control protocols for Agent 1, 2, and 3. } 
		\label{fig:7}
	\end{figure}

\begin{figure}[bt]
		\centering
		\includegraphics[width=10cm] {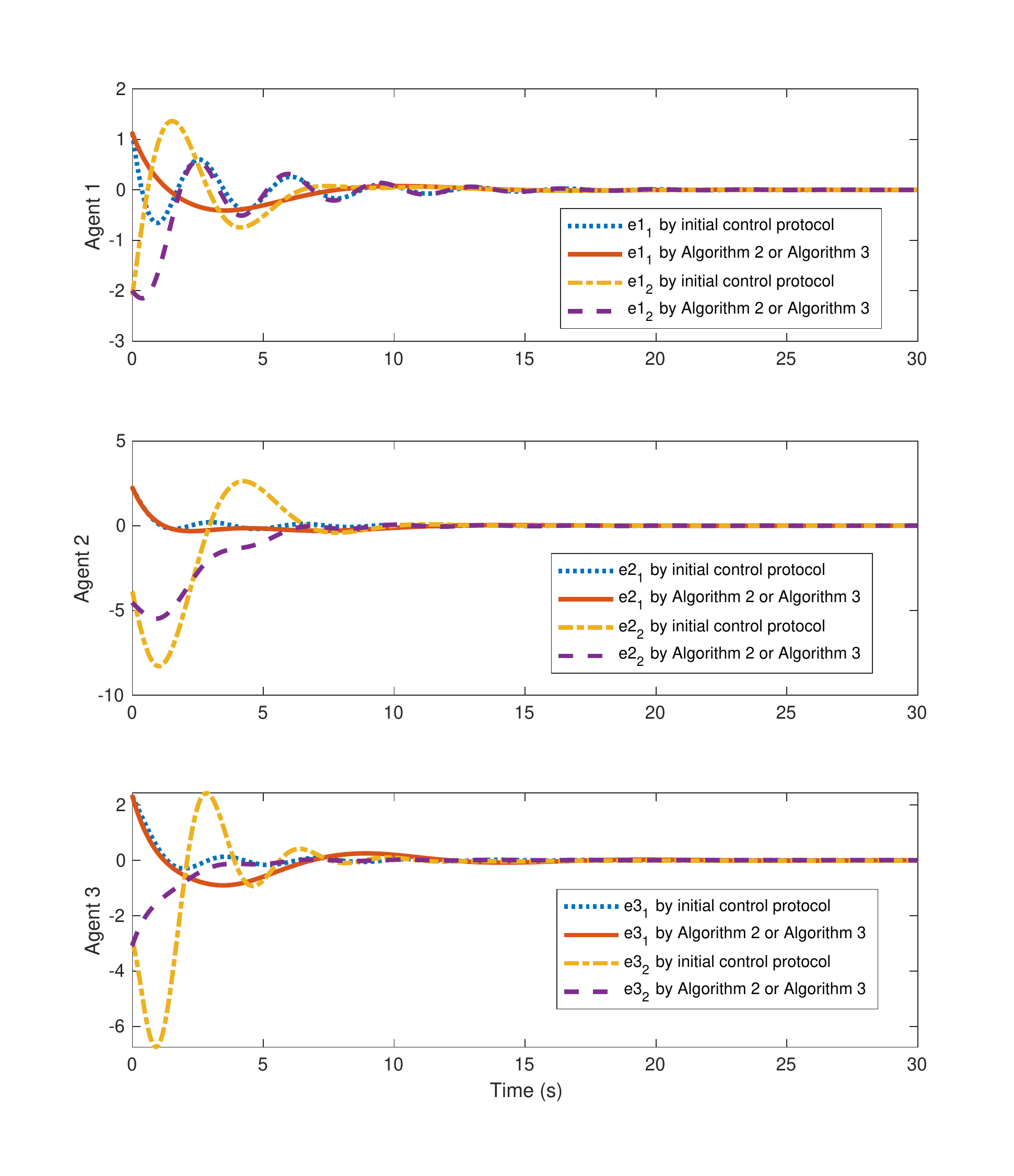}\\
		\caption{Comparisons of system performance under the initial and optimal control protocols for Agent 4 and 5.} 
		\label{fig:8}
	\end{figure}

	To compare performance of the developed Algorithm 1 with the initial control protocols, we let $w_0$ randomly generated around the origin with $w_0 \neq 0$ and set the same initial conditions for the five agents as 
	%
	\begin{align*}
		x_{10} & = \left[1.2,~-0.8,~0\right]^{\mathrm{T}}, \quad  x_{20}  = \left[ 1.6,~-0.5,~ 0 \right]^{\mathrm{T}}, \\
		x_{30} &  = \left[1.7,~ -0.4 ,~ 0 \right]^{\mathrm{T}}, \quad x_{40} = \left[0.8,~ -0.1,~ 0 \right]^{\mathrm{T}},  \\
		x_{50} & =  \left[  0.9,~ -0.4 ,~ 0 \right]^{\mathrm{T}},  \quad \xi_{10}  = \left[ 0.5,~ -0.4 \right]^{\mathrm{T}}, \\
		\xi_{20} & = \left[ 0.1,~ -0.2 \right]^{\mathrm{T}},  \quad \xi_{30}  = \left[ 0.1,~ -0.6 \right]^{\mathrm{T}}, \\
		\xi_{40}  & =  \left[ 0.3 ,~ -0.2 \right]^{\mathrm{T}},  \quad \xi_{50}  =  \left[ 0.3,~  -0.1\right]^{\mathrm{T}}, 
	\end{align*}
	Then we rerun the simulation program and comparisons of the tracking errors of the five agents are depicted in Figures~\ref{fig:7} and \ref{fig:8}. It can be seen that the proposed Algorithm 1, not only stabilize the system, but also make the MAS achieve optimality. This indicates that the proposed \emph{Problem 1} is solved. 
	

	\section{Conclusions}   \label{sec6}
	In this paper, we have proposed the online adaptive learning solutions for optimal synchronization control of the MASs with heterogeneous dynamics and leader disturbance. In Section \ref{sec3}, we design the distributed control protocol to make the output of each agent synchronize with the reference output. In Section \ref{sec4}, the optimal feedback gains of the distributed protocols are learned via PI technique by minimizing the output tracking error. Finally, in Section \ref{sec5}, the proposed approach is verified by an illustrative numerical example with six agents.

\bibliographystyle{unsrt}  
\bibliography{references}  

%
%
%
%

\end{document}